\documentstyle[aps,prd,floats,epsf]{revtex}
\tighten
\draft
\begin{document}

\twocolumn[\hsize\textwidth\columnwidth\hsize\csname
@twocolumnfalse\endcsname
\title{Mass signature of supernova $\nu_\mu$ and $\nu_\tau$ neutrinos
in SuperKamiokande}
\author{J.~F. Beacom\thanks{Electronic address:
        {\tt beacom@citnp.caltech.edu}} and
        P. Vogel\thanks{Electronic address:
        {\tt vogel@lamppost.caltech.edu}}}
\address{Department of Physics, California Institute of Technology\\
         Pasadena, CA 91125, USA}
\date{February 25, 1998}
\maketitle

\begin{abstract}

The $\nu_\mu$ and $\nu_\tau$ neutrinos (and their antiparticles) from
a Galactic core-collapse supernova can be observed in a
water-\v{C}erenkov detector by the neutral-current excitation of
$^{16}$O.  The number of events expected is several times
greater than from neutral-current scattering on electrons.  The
observation of this signal would be a strong test that these neutrinos
are produced in core-collapse supernovae, and with the right
characteristics.  In this paper, this signal is used as the basis for
a technique of neutrino mass determination from a future Galactic
supernova.  The masses of the $\nu_\mu$ and $\nu_\tau$ neutrinos can
either be measured or limited by their delay relative to the
$\bar{\nu}_e$ neutrinos.  By comparing to the high-statistics
$\bar{\nu}_e$ data instead of the theoretical expectation, much of the
model dependence is canceled.  Numerical results are presented for a
future supernova at 10 kpc as seen in the SuperKamiokande detector.
Under reasonable assumptions, and in the presence of the expected
counting statistics, $\nu_\mu$ and $\nu_\tau$ masses down to about 50
eV can be simply and robustly determined.  The signal used here is
more sensitive to small neutrino masses than the signal based on
neutrino-electron scattering.

\end{abstract}

\pacs{14.60.Pq, 97.60.Bw, 25.30.Pt, 95.55.Vj}

\vspace{0.5cm}]
\narrowtext

%%%%%%%%%%%%%%%%%%%%%%%%%%%%%%%%%%%%%%%%%%%%%%%%%%%%%%%%%%%%%%%%%%%%%%%%%%%%
%%%%%%%%%%%%%%%%%%%%%%%%%%%%%%%%%%%%%%%%%%%%%%%%%%%%%%%%%%%%%%%%%%%%%%%%%%%%

\section{Introduction}

When the core of a large star ($M \ge 8 M_{\odot}$) runs out of
nuclear fuel, it collapses and forms a proto-neutron star with a
central density well above the normal nuclear density (for a review of
type-II supernova theory, see Ref.~\cite{Bethe}). The total energy
released in the collapse, i.e., the gravitational binding energy of
the core ($E_B \sim G_N M_ {\odot}^2/R$ with $R \sim$ 10 km), is about
$3 \times 10^{53}$ ergs; about 99\% of that is carried away by
neutrinos and antineutrinos, the particles with the longest mean free
path.  The proto-neutron star is dense enough that neutrinos diffuse
outward over a timescale of several seconds, maintaining thermal
equilibrium with the matter.  When they are within about one mean free
path of the edge, they escape freely, with a thermal spectrum
characteristic of the surface of last scattering.  The luminosities of
the different neutrino flavors are approximately equal.

Those flavors which interact the most with the matter will decouple at
the largest radius and thus the lowest temperature.  The $\nu_\mu$ and
$\nu_\tau$ neutrinos and their antiparticles, which we collectively
call $\nu_x$ neutrinos, have only neutral-current interactions with
the matter, and therefore leave with the highest temperature, about 8
MeV (or $\langle E \rangle \simeq$ 25 MeV).  The $\bar{\nu}_e$ and
$\nu_e$ neutrinos have also charged-current interactions, and so leave
with lower temperatures, about 5 MeV ($\langle E \rangle \simeq$ 16
MeV) and 3.5 MeV ($\langle E \rangle \simeq$ 11 MeV), respectively.
The $\nu_e$ temperature is lower because the material is neutron-rich
and thus the $\nu_e$ interact more than the $\bar{\nu}_e$.  The
observation of supernova $\nu_x$ neutrinos would allow the details of
the picture above to be tested.  For a detailed description of the
supernova neutrino emission, including the justification of our choice
of temperatures, see Refs.~\cite{Woosley,Janka}.

Even after many decades of experiments, it is still not known whether
neutrinos have mass.  Results from several experiments strongly
suggest that neutrino flavor mixing occurs in solar, atmospheric, and
accelerator neutrinos, and proof of mixing would be a proof of mass.
The requirement that neutrinos do not overclose the universe gives a
bound for the sum of masses of stable neutrinos (see \cite{Raffelt}
and references therein):
\begin{equation}
\sum_{i=1}^3 m_{\nu_i} \lesssim 100 {\rm\ eV}\,.
\label{eq:cosmo}
\end{equation}
However, direct kinematic tests of neutrino mass currently give limits
for the masses compatible with the above cosmological bound only for
the electron neutrino, $m_{\bar{\nu}_e} \lesssim 5$ eV\cite{Belesev}.
For the $\nu_\mu$ and $\nu_\tau$ neutrinos the kinematic limits far
exceed the cosmological bound: $m_{\nu_\mu} < 170$ keV\cite{RPP}, and
$m_{\nu_\tau} < 24$ MeV\cite{RPP}.  It is very unlikely that direct
kinematic tests can improve these mass limits by the necessary orders
of magnitude any time soon.

As we will show in detail below, the most promising method for
determining these masses is with supernova neutrinos.  Even a tiny
mass will make the velocity slightly less than for a massless
neutrino, and over the large distance to a supernova will cause a
measurable delay in the arrival time.  A neutrino with a mass $m$ (in
eV) and energy $E$ (in MeV) will experience an energy-dependent delay
(in s) relative to a massless neutrino in traveling over a distance D
(in 10 kpc) of
\begin{equation}
\Delta t(E) = 0.515 \left(\frac{m}{E}\right)^2 D\,,
\label{eq:delay}
\end{equation}
where only the lowest order in the small mass has been kept.  Since
one expects one type-II supernova about every 30 years in our
Galaxy\cite{SNrate}, and since supernova neutrino detectors are
currently operating, it is worthwhile to consider whether mass limits
(or values) for $\nu_x$ compatible with the cosmological bound,
Eq.~(\ref{eq:cosmo}), can be obtained.

The problem of $\nu_x$ mass determination with supernova neutrinos
in existing
(e.g., Refs.~\cite{Wolfenstein,Seckel,Krauss,Burrows,Fiorentini})
and proposed detectors
(e.g., Refs.~\cite{Boron,SNBO,OMNIS})
has been considered before.  The present work differs from the
previous ones by the method with which the $\nu_x$ are detected:
inelastic scattering on $^{16}$O nuclei followed by proton or neutron
emission, and subsequent gamma decay of excited $^{15}$N or $^{15}$O
nuclei, as suggested in Ref.~\cite{LVK}.  We describe this signal and
its time structure in Section II. In Section III we discuss the most
relevant case of small masses.  We find the smallest $\nu_x$ mass that
is recognizably different from zero in the presence of the expected
finite counting statistics.  In Section IV we show that the mass range
is also limited from above.  If the $\nu_x$ mass is too large, the
signal is broadened to such a degree that it disappears into the
unavoidable background.  We find the largest detectable $\nu_x$ mass.
Finally, in Section V we summarize our findings.

%%%%%%%%%%%%%%%%%%%%%%%%%%%%%%%%%%%%%%%%%%%%%%%%%%%%%%%%%%%%%%%%%%%%%%%%%%%%
%%%%%%%%%%%%%%%%%%%%%%%%%%%%%%%%%%%%%%%%%%%%%%%%%%%%%%%%%%%%%%%%%%%%%%%%%%%%

\section{Characteristics of the model}

%%%%%%%%%%%%%%%%%%%%%%%%%%%%%%%%%%%%%%%%%%%%%%%%%%%%%%%%%%%%%%%%%%%%%%%%%%%%

\subsection{Neutrino scattering rate}

We assume that the double differential number distribution of
neutrinos of a given flavor (one of
$\nu_e,\bar{\nu}_e,\nu_\mu,\bar{\nu}_\mu,\nu_\tau,\bar{\nu}_\tau$) at
the source can be written in the product form:
\begin{equation}
\frac{d^2 N_\nu}{dE dt_i} = F(E)\,G(t_i)\,,
\end{equation}
where $E$ is the neutrino energy and $t_i$ is the time at the source.
The double integral of this quantity is the total number of emitted
neutrinos of that flavor $N_\nu$.  This form assumes that the energy
spectrum $F(E)$ is time-independent; the time dependence of the source
is parametrized solely by $G(t_i)$.  The reasons for assuming that the
energy and time dependences are separable will be given below.  The
most general form would allow $F = F(E,t_i)$, e.g., a time-dependent
temperature.  The luminosity is
\begin{equation}
L(t_i) = \int dE\, E\, \frac{d^2 N_\nu}{dE dt_i} 
= \langle E \rangle\, G(t_i) \int dE\, F(E)\,,
\end{equation}
where $\langle E \rangle$ is the (time-independent) average energy.
If the energy spectrum is normalized as
\begin{equation}
f(E) = \frac{F(E)}{\int dE\, F(E)}\,,
\end{equation}
then we can write
\begin{equation}
\frac{d^2 N_\nu}{dE dt_i} = f(E)\,\frac{L(t_i)}{\langle E \rangle}\,.
\label{eq:flux}
\end{equation}
This form is convenient since we assume, as stated earlier, that the
luminosities of the different flavors are approximately equal at every
time $t_i$.  The energy spectrum $f(E)$ will be taken to be thermal,
and the luminosity $L(t_i)$ will be taken to have a very sharp rise
and an exponential decline.  The arrival time of a neutrino of mass
$m$ at the detector is $t = t_i + D + \Delta t(E)$, where $D$ is the
distance to the source, and the energy-dependent time delay is given
by Eq.~(\ref{eq:delay}).  For convenience, we drop the constant $D$.
Then the double differential number distribution of neutrinos at the
detector is given by
\begin{eqnarray}
\frac{d^2 N_\nu}{dE dt} & = & 
\int dt_i\, \frac{d^2 N_\nu}{dE dt_i}
\delta(t - t_i - \Delta t(E)) \nonumber \\
& = & f(E)\,\frac{L(t - \Delta t(E))}{\langle E \rangle}\,.
\end{eqnarray}
Note that because of the mass effects, this is no longer the product
of a function of energy alone and a function of time alone.  The
number flux of neutrinos at the detector is obtained by dividing this
by $4\pi D^2$.  The scattering rate for a given neutrino reaction is
then
\begin{equation}
\frac{dN_{sc}}{dt} = N_{H_2 0}\;
 n \int dE\,\sigma(E) \frac{1}{4\pi D^2} \frac{d^2 N_\nu}{dE dt}\,,
\end{equation}
where $N_{H_2 O}$ is the number of water molecules in the detector,
$\sigma(E)$ the cross section for a neutrino of energy $E$ on the
target particle, and $n$ the number of targets per water molecule for
the given reaction.  Using the results above,
\begin{equation}
\frac{dN_{sc}}{dt} = N_{H_2 0} \frac{1}{4\pi D^2}
 \frac{1}{\langle E \rangle}\,
 n \int dE\,\sigma(E) f(E) L(t - \Delta t(E))\,.
\end{equation}
In more convenient units, the scattering rate (per s) is:
\begin{equation}
\frac{dN_{sc}}{dt} = C
\int dE\,f(E) \left[\frac{\sigma(E)}{10^{-42} {\rm cm}^2}\right]
\left[\frac{L(t - \Delta t(E))}{E_B/6}\right]\,,
\label{eq:rate}
\end{equation}
where
\begin{equation}
C = 9.21
\left[\frac{E_B}{10^{53} {\rm\ ergs}}\right]
\left[\frac{1 {\rm\ MeV}}{T}\right]
\left[\frac{10 {\rm\ kpc}}{D}\right]^2
\left[\frac{{\rm det.\ mass}}{1 {\rm\ kton}}\right]
\,n\,,
\label{eq:C}
\end{equation}
$T$ is the spectrum temperature (where we assume $\langle E \rangle =
3.15 T$, as appropriate for a Fermi-Dirac spectrum), and $f(E)$ is in
MeV$^{-1}$.  Since the luminosities are equal for each flavor, the
total binding energy released in a given flavor is $E_B/6$ (we ignore
the small effect associated with the neutronization burst).  When an
integral over all arrival times is made, the luminosity term in square
brackets integrates to one, giving for the total number of scattering
events:
\begin{equation}
N_{sc} = C
\int dE\,f(E) \left[\frac{\sigma(E)}{10^{-42} {\rm cm}^2}\right]\,.
\label{eq:total}
\end{equation}
The formulae in this section were derived for a nonzero neutrino mass;
for massless neutrinos, simply take $\Delta t(E) = 0$ throughout.  In
particular, note that in Eq.~(\ref{eq:rate}) the luminosity term then
can be taken outside of the integral, making the time dependence of
the scattering rate simply a constant times the time dependence of the
luminosity.

%%%%%%%%%%%%%%%%%%%%%%%%%%%%%%%%%%%%%%%%%%%%%%%%%%%%%%%%%%%%%%%%%%%%%%%%%%%%

\subsection{Details of the model}

As noted above, we assume that the energy distribution for a given
flavor of neutrinos is time-independent, e.g., that the temperature
does not vary with time.  While the temperature really will vary with
time, the variation is probably not large (see e.g., Fig.~3 of
Ref.~\cite{Woosley}, but note that those ``average'' energies are
defined as $\langle E^2 \rangle/\langle E \rangle$).  Also, recent
numerical models of supernovae disagree on the form of the variation,
and even whether it is rising or falling.  A well-motivated form for
temperature variation may eventually be obtained from the supernova
$\bar{\nu}_e$ data or from more-developed numerical models.  The
analysis of this paper could be easily modified to allow a varying
temperature; until there is a compelling reason to use a particular
form, we simply use a constant temperature.

The energy distribution is taken to be a Fermi-Dirac distribution,
characterized only by a temperature.  We take $T = 8$ MeV for $\nu_x$,
$T = 5$ MeV for $\bar{\nu}_e$, and $T = 3.5$ MeV for $\nu_e$.  These
temperatures are consistent with numerical models, e.g., in
Ref.~\cite{Janka}.  More elaborate models also introduce a chemical
potential parameter to reduce the high-energy tail of the Fermi-Dirac
distribution.  That reduces the number of scattering events, but makes
the dominant contribution to the cross section occur at a lower
neutrino energy, thus giving a larger delay.

Numerical supernova models suggest that the neutrino luminosity rises
quickly over a time of order 0.1 s, and then falls over a time of
order several seconds.  Therefore, the luminosity used in our
numerical simulation is composed of two pieces.  The first gives a
very short rise from zero to the full height over a time 0.09 s, using
one side of a Gaussian with $\sigma$ = 0.03 s.  The rise is so fast
that the details of its shape are irrelevant.  The second piece is an
exponential decay with time constant $\tau$ = 3 s.  The luminosity
then has a width of 10 s or so, consistent with the SN 1987A
observations.  The detailed form of the neutrino luminosity is less
important than the general shape features and their characteristic
durations.  In Ref.~\cite{Woosley}, the neutrino luminosity actually
decreases as a power law, and does so somewhat faster than our
exponential.  The slower the decay, the harder it is to see mass
effects, so our choice is actually somewhat conservative.

Throughout the paper, we assume that the distance to the supernova is
$D = 10$ kpc, approximately the distance to the Galactic center.

%%%%%%%%%%%%%%%%%%%%%%%%%%%%%%%%%%%%%%%%%%%%%%%%%%%%%%%%%%%%%%%%%%%%%%%%%%%%

\subsection{Characteristics of SuperKamiokande}

In this paper, all of the results are for the SuperKamiokande (SK)
detector.  The analysis here could be easily applied to any
water-\v{C}erenkov detector.  Its large size, low threshold, and low
background rate make it very well-suited to detect a Galactic
supernova.  We assume an energy threshold of 5 MeV; presently it is a
little bit higher but has been lowered a few times.  The full volume
of the main tank is 32 kton.  From SK conference talks\cite{SKbg}, we
estimate the time-independent background rate for the inner fiducial
22.5 kton volume to be about 0.1 s$^{-1}$ for a threshold of 5 MeV.
For the full 32 kton volume, we estimate that the background rate can
be no more than several times worse than 0.1 s$^{-1}$, again for a
threshold of 5 MeV.  For the low-mass search in Section III, we assume
that the full 32 kton volume is used.  The exact value of the
time-independent background rate is completely irrelevant in that
search.  For the high-mass search in Section IV, we assume that only
the inner 22.5 kton will be used, since in that case the
time-independent background rate would be an important factor.  Using
only the inner volume will decrease the number of signal events by a
factor 1.4, while decreasing the background by a factor of at least a
few.

%%%%%%%%%%%%%%%%%%%%%%%%%%%%%%%%%%%%%%%%%%%%%%%%%%%%%%%%%%%%%%%%%%%%%%%%%%%%

\subsection{Description of the signal}

The cross section for the neutral-current excitation of $^{16}$O by
neutrinos was computed numerically in Ref.~\cite{LVK}.  It was assumed
to be a two-step process, of excitation of $^{16}$O to the continuum,
followed by decays into various final states.  The principal branches
in this decay are to states of $^{15}{\rm N + p}$ and $^{15}{\rm O +
n}$.  For $\nu_x$ neutrinos with a thermal spectrum with $T = 8$ MeV,
the combined branching ratio for these final states is about 95\%.  If
the decay is to a bound excited state of the daughter nucleus, then
the daughter will decay by gamma emission.  At the relevant excitation
energies in $^{16}$O, the branching ratio to these states in the
daughters is about 30\%.  The crucial point is that in both $^{15}$N
and $^{15}$O all gamma rays lie between 5 and 10 MeV and can thus be
detected in SK.  The other 70\% of the branching ratio involves decays
to the ground state of the daughters without gamma emission.  In order
to get to a final state with a gamma, the neutrino energy must be
greater than about 20 MeV.  Because of this high threshold, and
because of the lower $\nu_e$ and $\bar{\nu}_e$ temperatures, these
reactions contribute only at the 2\% level compared to the $\nu_x$
reactions, and hence are ignored\cite{LVK}.

In Refs.~\cite{LVK,Kolbe}, the neutral-current cross sections were
calculated numerically and folded with thermal neutrino spectra of
different temperatures.  For the present purpose we need the cross
section for a given neutrino energy.  It turns out that the simple
parameterized form $\sigma(E) = \sigma_0 (E - 15)^4$, with the
neutrino energy $E$ in MeV and $\sigma_0 = 0.75 \times 10^{-47}$
cm$^2$ describes quite well the cross section for a neutrino to excite
$^{16}$O.  In the fit we assumed that the branching ratio for states
that end with gamma emission is independent of neutrino energy.  All
such branches are included in this cross section above, and we have
summed the cross sections for neutrinos and antineutrinos (for just
one flavor), as well as both final channels.  The fit values agree
with the numerical calculations at the 10\% level over four orders of
magnitude in the thermally-averaged excitation cross section.  This
fit will certainly not hold at higher energies which are however
irrelevant in the present context.

In order to estimate the delay, Eq.~(\ref{eq:delay}) can be evaluated
with a typical neutrino energy.  However, one should not use the
average energy, $\langle E \rangle = 25$ MeV.  Rather, one should use
the energy for which $f(E)\sigma(E)$ peaks.  For this reaction, this
``Gamow peak'' energy is $E \approx 60$ MeV, i.e., considerably larger
than $\langle E \rangle$.  The fact that the neutrinos have a spectrum
of energies means that different values of $E$ contribute to the time
delay, causing dispersion of the neutrino pulse as it travels from the
supernova.  It turns out that for the small masses we are primarily
interested in these dispersive effects are minimal.

The signal associated with the gamma emission described above will not
be the dominant signal of a Galactic supernova in SuperKamiokande.
Rather, the dominant events will be the positrons from $\bar{\nu}_e +
p \rightarrow e^+ + n$, which give a smooth continuum in positron
energy, peaking at about 20 MeV.  The expected numbers of events for
various reactions were calculated with Eq.~(\ref{eq:total}) and are
given in Table I.  For the $\bar{\nu}_e$ absorption on proton
reaction, recoil and weak magnetism effects were taken into account,
which slightly reduces the cross section.  There are also
charged-current reactions on $^{16}$O\cite{Haxton}; these increase the
dominant positron signal by about 1\%.  Since events from the
electron-scattering channels are forward-peaked, we assume that they
are removed by an angular cut.  Therefore, in our analysis we use only
the events from $\bar{\nu}_e$ absorption on protons and the $\nu_x$
excitation of $^{16}$O.

The gammas from the neutral-current reactions above are at several
discrete energies ranging from 5.2 MeV to 9.9 MeV.  These are subject
to some smearing, due to the finite resolution, giving few narrow
peaks on top of the smooth distribution of positrons as shown in
Fig.~2 of Ref.~\cite{LVK}.  For simplicity, we treat the energy range
from threshold to 10 MeV as one bin, and assume that losses due to the
threshold or efficiency are minimal.

In Ref.~\cite{LVK} numbers of events from different reactions were
calculated relative to each other, with the overall scale set by the
total number of $\bar{\nu}_e$ events from Ref.~\cite{Totsuka}.
However, the number of $\bar{\nu}_e$ events corresponding to $T = 3$
MeV from Ref.~\cite{Totsuka} was used.  This was not really
consistent, and would not be consistent here either, since for the
$\bar{\nu}_e$ neutrinos, $T = 5$ MeV is assumed here and in
Ref.~\cite{LVK}.  Consequently, we use instead Eq.~(\ref{eq:total}) to
calculate the number of events for $T = 5$ MeV.  We verified that the
rates based on Eq.~(\ref{eq:total}) agree with the numbers given in
Ref.~\cite{Totsuka} when consistent temperatures are used.  Note that
the results of Ref.~\cite{LVK} are changed only by increasing the
number of events in each reaction by a factor of about 2.

\begin{table}[t]
\caption{Calculated numbers of events expected in SK with a 5 MeV
threshold and a supernova at 10 kpc.  The other parameters (e.g.,
neutrino spectrum temperatures) are given in the text.  In rows with
two reactions listed, the number of events is the total for both.  The
second row is a subset of the first row that is an irreducible
background to the reactions in the third and fourth rows.}
\begin{tabular}{|l|l|}
\hline
reaction & number of events \\
\hline\hline
$\bar{\nu}_e + p \rightarrow e^+ + n$ & 8300 \\
\hline
$\bar{\nu}_e + p \rightarrow e^+ + n$
 \phantom{aa} ($E_{e^+} \leq 10$ MeV) & 530 \\
\hline
$\nu_\mu + ^{16}{\rm O} \rightarrow \nu_\mu + \gamma + X$ & 355 \\
$\bar{\nu}_\mu + ^{16}{\rm O} \rightarrow \bar{\nu}_\mu + \gamma + X$ & \\
\hline
$\nu_\tau + ^{16}{\rm O} \rightarrow \nu_\tau + \gamma + X$ & 355 \\
$\bar{\nu}_\tau + ^{16}{\rm O} \rightarrow \bar{\nu}_\tau + \gamma + X$ & \\
\hline
$\nu_e + e^- \rightarrow \nu_e + e^-$ & 200 \\
$\bar{\nu}_e + e^- \rightarrow \bar{\nu}_e + e^-$ & \\
\hline
$\nu_\mu + e^- \rightarrow \nu_\mu + e^-$ & 60 \\
$\bar{\nu}_\mu + e^- \rightarrow \bar{\nu}_\mu + e^-$ & \\
\hline
$\nu_\tau + e^- \rightarrow \nu_\tau + e^-$ & 60 \\
$\bar{\nu}_\tau + e^- \rightarrow \bar{\nu}_\tau + e^-$ & \\
\end{tabular}
\end{table}

%%%%%%%%%%%%%%%%%%%%%%%%%%%%%%%%%%%%%%%%%%%%%%%%%%%%%%%%%%%%%%%%%%%%%%%%%%%%
%%%%%%%%%%%%%%%%%%%%%%%%%%%%%%%%%%%%%%%%%%%%%%%%%%%%%%%%%%%%%%%%%%%%%%%%%%%%

\section{Low-mass case}

In this section, we detail the strategy used in the analysis.  First,
the $\bar{\nu}_e$ mass is low enough that it can be neglected.  Our
final result is that one can reach sensitivity down to a $\nu_x$ mass
of about 50 eV.  Since the $\bar{\nu}_e$ mass is at least 10 times
smaller, and since the delay depends quadratically on the mass, this
neglect is justified.  This establishes the key point of our
technique: that we can use the $\bar{\nu}_e$ events as a clock by
which to measure the possible delay of the $\nu_x$ neutrinos.  Under
our assumption that the temperatures are approximately constant, the
only time dependence of the $\bar{\nu}_e$ scattering rate is from the
$\bar{\nu}_e$ luminosity itself (see Eq.~(\ref{eq:rate}) with $m =
0$).  In contrast, the time dependence of the $\nu_x$ scattering rate
is determined both by the $\nu_x$ luminosity and the delaying effects
of a possible mass.  Thus the effects of a mass can be tested for by
comparing the scattering rates of the $\bar{\nu}_e$ and $\nu_x$ events
as a function of time.  In other words, we are looking for time
dependence in the $\nu_x$ rate beyond that expected from the
luminosity variation alone.  In order to implement this, we define two
rates, as follows.

The scattering rate of $\bar{\nu}_e$ events with $E_{e^+} > 10$ MeV
will be called the Reference $R(t)$.  This contains $\approx 8300 -
530 \approx 7800$ events.  The time dependence of $R(t)$ is completely
determined by the time dependence of the luminosity.  Its shape is
generic for massless neutrinos.  The Signal $S(t)$ has three
components.  The first is the scattering rate for the 355 events from
the combined $\nu_\mu$ and $\bar{\nu}_\mu$ on $^{16}$O reactions.  The
second is the same for the 355 combined $\nu_\tau$ and
$\bar{\nu}_\tau$ events.  The third is the scattering rate for the 530
$\bar{\nu}_e$ events with $E_{e^+} < 10$ MeV.  We will assume that
some portion of the Signal $S(t)$ events are massive (either all
$\nu_\tau$ events or all $\nu_\mu$ and $\nu_\tau$ events).  All of the
other events in $S(t)$ are then massless background events.  Because
some of the $S(t)$ events will be massive, the shape of $S(t)$ will be
distorted.  In particular, it will be delayed and broadened.

In a given experiment (i.e., one supernova), the Signal $S(t)$ and the
Reference $R(t)$ will be measured.  In order to facilitate comparison
of their shapes, the curve $R(t)$ can be scaled down to the number of
events in $S(t)$.  The curve $S(t)$ shows how the data look, with a
possible unknown $\nu_\tau$ mass, and the curve $R(t)$ shows how they
would look if all of the events were massless.  The rates are shown in
Fig.~1, which depicts $S(t)$ under different assumptions about the
$\nu_\tau$ mass.  The shape of $R(t)$ is the same as that of $S(t)$
when $m_{\nu_\tau} = 0$.  The curve $R(t)$ will be measured, and so
will be there to compare the measured $S(t)$ to.  As the $\nu_\tau$
mass is increased, the delayed $\nu_\tau$ events separate from the
massless events more and more.  For $m = 125$ eV, the scattering rate
over the first 1 s or so is just that from the remaining massless
events.  The effect of a mass is to diminish the rate at early times
and enhance it at late times (since the normalization is preserved,
these are roughly equivalent statements).

In a real experiment, statistical fluctuations will mask the effect of
a mass.  The Reference $R(t)$ contains approximately 7800 events, and
thus has small relative fluctuations.  The Signal $S(t)$ contains
approximately 1240 events, and therefore has larger relative
fluctuations.  Each of those curves is subject to fluctuations in the
total number of events as well as fluctuations in any time interval.
Consider for a moment the events in the first 1 s of Fig.~1.  There
are 336 events expected in the $m_{\nu_\tau} = 0$ eV case, and 302
events expected in the $m_{\nu_\tau} = 50$ eV case.  As noted, the
Reference has smaller fluctuations, so for now take the total 336 as
exact.  The counting error on the Signal in that interval will be of
order $\sqrt{302} \approx 17$.  If the number of events in this bin
fluctuates up by about two sigma, then the number of events in the
Signal over this interval would match the number expected for the
massless case, and we would have to conclude that most probably the
mass of the $\nu_\tau$ is zero.  In the analysis below, we use much
more of the data, but the idea is the same: it is possible for one
mass case to fake another through fluctuations.  The degree to which
this can occur depends primarily on the number of events expected in
the Signal.  We will restrict the range of fluctuations that we
consider to be likely by choosing confidence levels.

To treat the expected fluctuations properly, we use a Monte Carlo
technique to generate representative statistical instances of the
theoretical forms for $R(t)$ and $S(t)$.  Each run represents one
supernova as seen in SK.  The total number of events expected in
$R(t)$ is known.  In each particular run, this total is subject to
Poisson fluctuations.  We model this by picking a Poisson random
number from a distribution with mean given by the expected number of
events.  This gives the number of events for this particular run.  We
then use an acceptance-rejection method to sample the form $R(t)$
until the right number of events for that run is obtained.  This gives
a statistical instance of $R(t)$, typical of what might be seen in a
single experiment.  Then an exactly analogous technique is used to
generate the total number of events in $S(t)$ and a statistical
instance of the curve $S(t)$ itself.  The massless and massive
components of $S(t)$ are sampled separately, and are then added
together.

\begin{figure}[t]
\epsfxsize=3.25in \epsfbox{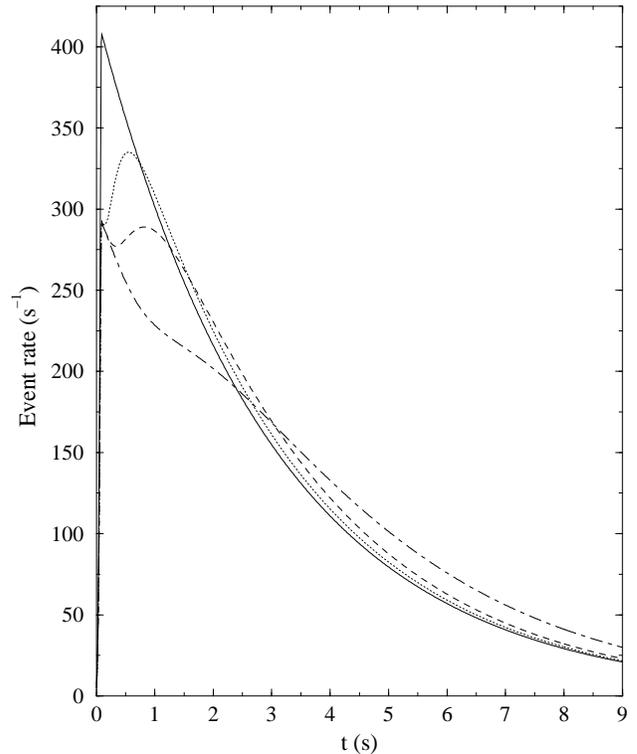}
\caption{The expected event rate in the absence of fluctuations for
the signal $S(t)$ is shown for different $\nu_\tau$ masses, as
follows: solid line, 0 eV; dotted line, 50 eV; dashed line, 75 eV;
dot-dashed line, 125 eV.  Of 1240 total events, 530 are massless
$\bar{\nu}_e$, 355 are massless $\nu_\mu$ and $\bar{\nu}_\mu$ , and
355 are massive $\nu_\tau$ and $\bar{\nu}_\tau$.  These totals count
events at all times; in the figure, only those with $t \le 9$ s are
shown.}
\end{figure}

One comment on the method of sampling is necessary.  No matter how the
generated rates are binned in time, this method ensures that in each
bin there are the correct Poisson fluctuations around the expected
number in that bin.  Therefore, this technique is equivalent to the
sometimes-seen technique of first establishing bins and the expected
number in each bin, and then picking a representative number of events
for that bin according to the appropriate Poisson distribution.
However, our method of generating representative Signal and Reference
data sets does not require binning.  If analysis of these data sets
uses bins, the bin size can be changed without regenerating the data.

Both of the tests developed below depend upon the shape of $S(t)$, and
not directly on the number of counts.  Direct tests for an excess or
deficit of counts are much more dependent on theory; this dependence
is largely canceled in our approaches.

%%%%%%%%%%%%%%%%%%%%%%%%%%%%%%%%%%%%%%%%%%%%%%%%%%%%%%%%%%%%%%%%%%%%%%%%%%%%

\subsection{$\chi^2$ analysis}

As discussed above, the presence of a mass in the Signal $S(t)$ will
cause its relative decrease at early times and relative increase at
late times in comparison with the Reference $R(t)$.  Whether or not it
can be seen is a question of the statistics of the event rates.  As a
first test we look for a shape distortion in $S(t)$ relative to $R(t)$
by making a $\chi^2$ test.  If the $\chi^2$ per degree of freedom
(d.o.f.) is of order unity, then the two curves are compatible at the
level of the errors, and there is no reason to invoke a mass.  If the
$\chi^2$/d.o.f.\ is large, then the two functions are incompatible,
which we take as evidence for a mass.  That is, we assume that there
are no other systematic effects which would give a large
$\chi^2$/d.o.f.; one always has to make some such assumption.

We first scale the Reference down to the number of events observed in
the Signal over the range $0 \le t \le t_{max}$.  As required for a
$\chi^2$ test, both sets are then binned so the continuous functions
$R(t)$ and $S(t)$ are replaced by discrete representations.  Bins of
constant width $\delta t$ are used.  The scaling is given by
\begin{equation}
\widetilde{R}_j = R_j
\frac{\sum^{N_{bin}}_{j = 1} S_j \delta t}
{\sum^{N_{bin}}_{j = 1} R_j \delta t}\,,
\end{equation}
where $t_{max} = N_{bin} \delta t$.  When $m = 0$, $\widetilde{R}_j =
S_j$, up to statistical fluctuations.  The $\chi^2$ is formed as
follows:
\begin{equation}
\chi^2/{\rm d.o.f.} = \frac{1}{N_{bin} - 1} 
\sum^{N_{bin}}_{j = 1}
\frac{(\widetilde{R}_j \delta t - S_j \delta t)^2}
{s {\widetilde{R}_j \delta t + S_j \delta t}}\,,
\end{equation}
where $j$ runs over the bins used.  The number of degrees of freedom
is reduced by one because we have normalized the Reference to the
Signal.  The factor $s$ is the ratio of the total numbers of events
(in $0 \le t \le t_{max}$) in the Signal and the Reference, and is
computed for each run in the Monte Carlo.  Even though
$\widetilde{R}_j \delta t \approx S_j \delta t$, the fluctuations in
$\widetilde{R}_j \delta t$ are much smaller since $\widetilde{R}_j
\delta t$ is scaled down from $R_j \delta t$, which has high
statistics.

It is important to stress that it is not enough to evaluate the
$\chi^2$ using the predicted curves for $R(t)$ and $S(t)$ based on the
analytic forms constructed with Eq.~(\ref{eq:rate}).  Doing so
neglects fluctuations, and always underestimates the $\chi^2$,
particularly near the small-mass limit that we are interested in
(since in the massless case $\widetilde{R}(t) = S(t)$).  Roughly
speaking, using the exact functions themselves in the $\chi^2$
underestimates the $\chi^2$/d.o.f.\ by about unity, and of course does
not give the error.  As explained above, we use the Monte Carlo
technique which properly treats statistical fluctuations, and leads to
a more conservative (and correct) mass limit.

Only a finite range of times was used in forming the $\chi^2$.  The
beginning of the first bin is taken to be where the events start.
With some 9000 total events expected, and a risetime of order 0.1 s,
the starting time can be reasonably well-defined.  In the Monte Carlo,
the starting time was held fixed (and not adjusted from the data on
each run).  The definition used amounts to calling the starting time
that point at which the $\bar{\nu}_e$ rate is about 1\% of its peak
rate.  The size of any ambiguity in the starting time is much smaller
than the bin size (discussed below), and so is regarded as irrelevant.

The ending time and the bin size must be chosen more carefully.  The
primary consideration is to maximize the extraction of the mass effect
in the presence of the statistical fluctuations.  Further, this must
be optimized for the case of a small mass (other cases are discussed
below).  In Fig.~1, one can see that for a given $m_{\nu_\tau}$, the
Signal $S(t)$ rejoins the Reference $R(t)$ at very late times (even
beyond the edge of the figure for the larger masses).  Once this has
happened, there is no benefit to going to larger times; in fact, one
only includes more statistical noise by doing so.  In the Monte Carlo
studies, it was found that $t_{max} = 9$ s and a bin size of $\delta t
= 1$ s were good choices (the final results are only weakly dependent
on these).  These are also very reasonable from a physical point of
view.  These have to be held fixed for all of the Monte Carlo runs,
since one cannot adjust these to a particular data set without
introducing bias.  These choices also ensure that we can completely
neglect the time-independent background rate of at most a few times
0.1 s$^{-1}$.

With these choices, one has a reasonable number (namely 8) of degrees
of freedom in the $\chi^2$, and a large number of events expected in
each bin.  The latter ensures that the Poisson errors on the counts in
each bin really are approximately Gaussian, as required in the
$\chi^2$ definition.  Up to fluctuations, the late-time bins all have
an excess.  Combining them would enhance the significance of this
excess, whereas for random fluctuations combining bins does not change
the significance.  The same is true for the early-time deficits.
However, one does not in general know where the transition point is
between these two regions; that is determined by the unknown mass.
The transition point cannot be determined from the data without
introducing bias.  Also, with too few bins, one does not satisfy the
requirements for defining a $\chi^2$ test.

Using the above procedure for analyzing each run (and in particular,
normalizing $R(t)$ to $S(t)$ over $0 \le t \le t_{max}$), we used the
Monte Carlo to simulate the results from $10^4$ supernovae.  For each
run, the $\chi^2$ analysis was performed.  For each fixed mass, a
variety of $\chi^2$ values are obtained, due to the finite statistics
in the Reference and the Signal.  These results were histogrammed as
$\chi^2$/d.o.f.  The relative frequencies of different
$\chi^2$/d.o.f.\ values are shown in the upper panel of Fig.~2 for a
few representative masses.  (Note that the number of Monte Carlo runs
determines only how smoothly these distributions are filled out; their
shape and placement is determined by the physics.)  For $m = 0$, the
resulting $\chi^2$/d.o.f.\ values of course fill out the usual
$\chi^2$/d.o.f.\ distribution with 8 degrees of freedom.

\begin{figure}[t]
\epsfxsize=3.25in \epsfbox{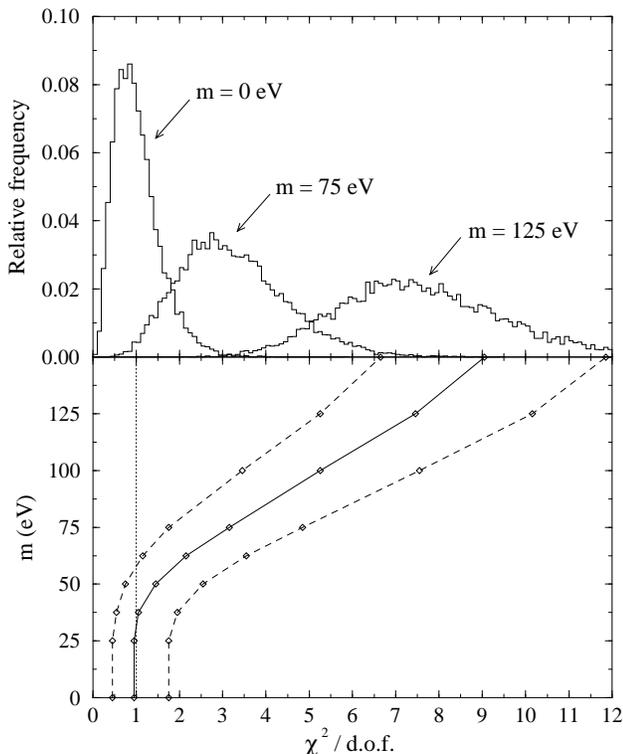}
\caption{The results of the $\chi^2$ analysis for a massive
$\nu_\tau$.  In the upper panel, the relative frequencies of various
$\chi^2$/d.o.f.\ values are shown for a few example masses.  In the
lower panel, the range of masses corresponding to a given
$\chi^2$/d.o.f.\ is shown.  The solid line is the 50\% confidence
level, and the upper and lower dashed lines are the 10\% and 90\%
confidence levels, respectively.  In this figure, $t_{max} = 9$ s, the
bin size used in the $\chi^2$ is $\delta t = 1$ s, and the time
constant of the exponential luminosity is $\tau = 3$ s.}
\end{figure}

These distributions are characterized by their central point and their
(asymmetric) width, using the 10\%, 50\%, and 90\% confidence levels.
That is, for each mass we determined the value of $\chi^2$/d.o.f.\
such that a given percentage of the Monte Carlo runs yielded a value
of $\chi^2$/d.o.f.\ less than that value.  With those three numbers,
we can characterize the results of complete runs with many masses much
more compactly, as shown in the lower panel.  For convenience, the
axes are inverted from how the plot was actually constructed.  That
is, given the $\chi^2$/d.o.f.\ which will be experimentally
determined, one can read off the range of masses that could have
likely given such a $\chi^2$/d.o.f.\ at these confidence levels.

%%%%%%%%%%%%%%%%%%%%%%%%%%%%%%%%%%%%%%%%%%%%%%%%%%%%%%%%%%%%%%%%%%%%%%%%%%%%

\subsection{$\langle t \rangle$ analysis}

The $\chi^2$ test above has the nice feature that it is a shape test,
and depends on the number of events only through the fluctuations.
One disadvantage is its dependence on binning, which obscures changes
over timescales smaller than of order the bin width, i.e., the effects
of sufficiently small masses.  Another is that the mass effect is not
always in the same sense.  At early times there is a deficit of
events, whereas at late times there is an excess; the $\chi^2$ is
insensitive to the difference between this distinctive feature and
random fluctuations of similar magnitude.  To get around these
problems, we introduce here tests of integral moments.  These do not
involve any binning.  The most basic effect of a mass is a delay; the
average arrival time always increases.  The test is simple,
intuitively obvious, and the effect is always in the same sense (up to
statistical fluctuations).  A mathematically analogous moments
analysis was made for electron recoil energies in the context of solar
neutrino oscillations in Ref.~\cite{Bahcall}.

Given the Reference $R(t)$, the average arrival time is defined as
\begin{equation}
\langle t \rangle_R = \frac{\sum_k t_k}{\sum_k 1} =
\frac{\int_0^{t_{max}} dt\,t R(t)}{\int_0^{t_{max}} dt\, R(t)} \,.
\end{equation}
The summation form is used for the Monte Carlo generated data sets,
where the sum is over events (not time bins) in the Reference with $0
\le t \le t_{max}$.  The integral form would be used if the
theoretical forms for the rates were given.  It is no longer necessary
to normalize the Reference to the Signal.  As with the $\chi^2$ test,
the starting time is assumed to be well-defined.  The choice of
$t_{max}$ follows from similar considerations as before.  The effect
of the finite number of counts in $R(t)$ is to give $\langle t
\rangle_R$ a statistical error.  This error is the intrinsic width of
the $R(t)$ distribution divided by the square root of the number of
events in the Reference.  Both the intrinsic width and number of
events depend on the choice of $t_{max}$.  By choosing a moderate
$t_{max}$, the intrinsic width of $R(t)$ can be restricted even while
most events are included.

Given the Signal $S(t)$, the average arrival time is defined similarly
as
\begin{equation}
\langle t \rangle_S = \frac{\sum_k t_k}{\sum_k 1} =
\frac{\int_0^{t_{max}} dt\,t S(t)}{\int_0^{t_{max}} dt\, S(t)} \,,
\end{equation}
where naturally the sums are now over events in the Signal.  While the
intrinsic widths of $R(t)$ and $S(t)$ are similar, the statistical
error on $\langle t \rangle_S$ is larger by factor of a few since
there are several times fewer events.  The effect of the mass is to
make $\langle t \rangle_S$ larger, i.e., to cause a delay.  (The mass
increases the intrinsic width of $S(t)$ only slightly.)

In order to cancel some systematic effects, we consider not $\langle t
\rangle_S$ as compared to theory but the difference $\langle t
\rangle_S - \langle t \rangle_R$ determined from the data.  The signal
of a mass is that this is greater than zero with statistical
significance.  From the Monte Carlo studies, $t_{max} = 9$ s was found
to be a very reasonable choice; most of the data are then included
while the range is kept small.  For this $t_{max}$, the
time-independent background events are negligible.  Again, while these
choices are somewhat optimal, the final results are not strongly
dependent on the particular values used as long as they are
reasonable.  Although the values of $\langle t \rangle$ depend on
$t_{max}$, the dependence is not strong.  For $t_{max} = 9$ s, a
change of 0.1 s in $t_{max}$ gives a change of about 0.01 s in
$\langle t \rangle$.  Note that any shift in the starting time will
cancel in the difference $\langle t \rangle_S - \langle t \rangle_R$
(as long as it doesn't change the numbers of events included).

Using the above procedure for analyzing a particular run, we again
used the Monte Carlo to simulate the results from $10^4$ supernovae.
Basically, things were done as above for $\chi^2$.  For each run,
$\langle t \rangle_S - \langle t \rangle_R$ was calculated and its
value histogrammed.  These distributions are again characterized by
their central point and their width, using the 10\%, 50\% (now also
the average), and 90\% confidence levels.  That is, for each mass we
determined the values of $\langle t \rangle_S - \langle t \rangle_R$
such that a given percentage of the Monte Carlo runs yielded a value
of $\langle t \rangle_S - \langle t \rangle_R$ less than that value.
Since these distributions are Gaussians, other confidence levels can
easily be constructed.  The results of this analysis are shown in
Fig.~3, which is analogous to Fig.~2.

For $t_{max} = 9$ s, $\langle t \rangle_R = 2.57$ s.  For larger
$t_{max}$, $\langle t \rangle_R$ tends to about 3 s, the value of the
exponential time constant in the luminosity.  The value of $\langle t
\rangle_S$ is of course larger by the mass effect.  As noted, the
error on each moment is the intrinsic width divided by the square root
of the number of events.  The intrinsic widths of the $R(t)$ and
$S(t)$ distributions are each of order a few seconds.  The numbers of
events are of order 8000 and 1200, respectively.  Note that the errors
on $\langle t \rangle_R$ and $\langle t \rangle_S$ are uncorrelated.

We also investigated the dispersion of the event rate in time as a
measure of the mass.  As noted above, a mass alone causes a delay, but
a mass and an energy spectrum also cause dispersion.  We defined the
dispersion as $\sqrt{\langle t^2 \rangle - \langle t \rangle^2}$,
where all integrals are as above defined up to $t_{max}$.  We found
that the effects were not statistically significant until the mass was
of order 150 eV or so; at such a large mass the statistical
significance of the change in $\langle t \rangle$ cannot be missed.

%%%%%%%%%%%%%%%%%%%%%%%%%%%%%%%%%%%%%%%%%%%%%%%%%%%%%%%%%%%%%%%%%%%%%%%%%%%%

\subsection{Comparison of techniques}

The analysis techniques presented above are appropriate for the case
in which the mass is either small or zero.  In this case, the signal
$S(t)$ and the reference $R(t)$ are not easily distinguished for
finite statistics.  Both the $\chi^2$ and $\langle t \rangle$ analyses
were optimized for this case by choosing a moderate $t_{max} = 9$ s,
which also allowed us to neglect the time-independent background.  In
each case, sensitivity to a $\nu_\tau$ mass of about 50 eV was found.
This is essentially the mass which cannot be missed even if there are
unfavorable statistical fluctuations.  Since the mass effects grow
quadratically, for larger masses the statistical significance of the
mass effects would be huge.

\begin{figure}[t]
\epsfxsize=3.25in \epsfbox{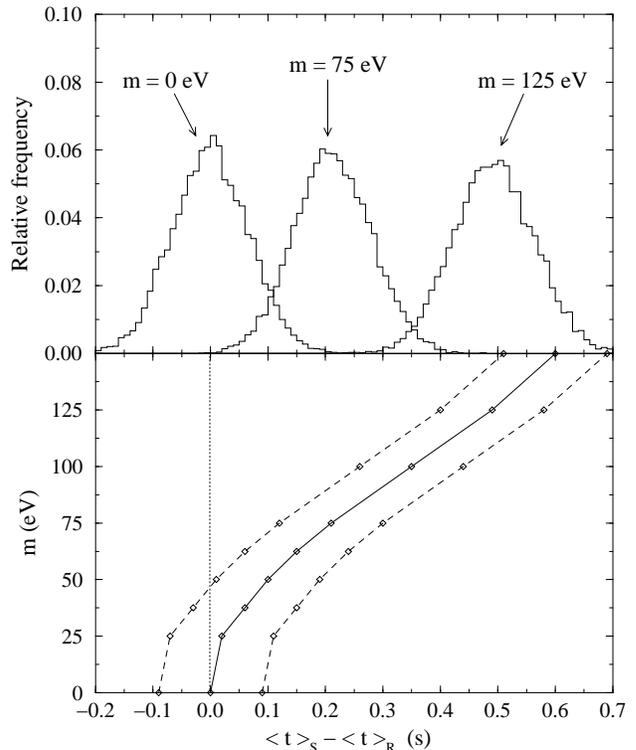}
\caption{The results of the $\langle t \rangle$ analysis for a massive
$\nu_\tau$.  In the upper panel, the relative frequencies of various
$\langle t \rangle_S - \langle t \rangle_R$ values are shown for a few
example masses.  In the lower panel, the range of masses corresponding
to a given $\langle t \rangle_S - \langle t \rangle_R$ is shown.  The
solid line is the 50\% confidence level, and the upper and lower
dashed lines are the 10\% and 90\% confidence levels, respectively.
In this figure, $t_{max} = 9$ s and the time constant of the
exponential luminosity is $\tau = 3$ s.}
\end{figure}

At a given mass, the ranges of $\chi^2$ or $\langle t \rangle_S -
\langle t \rangle_R$ values shown in the figures are the ranges of
probable values that would be seen in one experiment (i.e., one
supernova).  Those ranges are the result of properly taking into
account the expected statistical fluctuations of the Reference and
Signal (while the Signal error dominates, the Reference error was
included in the calculations).  For a given experiment, the values of
$\chi^2$ and $\langle t \rangle_S - \langle t \rangle_R$ can be
computed from the data.  The statistical errors on those quantities
can also be estimated from the data, and should be similar to what is
shown in the figures.

The results from both analysis techniques are essentially similar.
That is, the final results are not strongly dependent on the
statistical technique used, which is crucial.  Of course, the results
from the $\langle t \rangle$ analysis are slightly better, for the
reasons explained above.  The final figures for the $\langle t
\rangle$ analysis also allow other confidence levels to be constructed
easily.  Our $\chi^2$ test was designed

\newpage
\widetext

\begin{table}[t]
\caption{The results of the $\chi^2$ analysis for different cases.
The decay constant of the exponential luminosity is denoted by $\tau$.
If the masses are zero, the most probable $\chi^2/{\rm d.o.f.} = 1$.
For this $\chi^2$/d.o.f., the allowed mass ranges are given in the
second column; the lower limit of zero is the most probable mass, and
the upper limit is excluded at the 90\% confidence level.  The
smallest value of $\chi^2$/d.o.f.\ not compatible with $m = 0$ is
$\chi^2/{\rm d.o.f.} = 1.7$.  The corresponding allowed mass ranges
are given in the third column; both the upper and lower limits are
excluded at the 90\% confidence level.  The most probable mass is
given in parentheses.}
\begin{tabular}{|l|l|l|}
\hline
case & result for $\chi^2/{\rm d.o.f.} = 1$ &
result for $\chi^2/{\rm d.o.f.} = 1.7$ \\
\hline\hline
$\tau = 3$s; $m_{\nu_\mu} = 0,\, m_{\nu_\tau} = m$ & $0 \le m < 60$ eV 
 & $0 < m < 75$ eV\ \ ($m \simeq 55$ eV)\\
\hline
$\tau = 3$s; $m_{\nu_\mu} = m_{\nu_\tau} = m$ & $0 \le m < 40$ eV 
 & $0 < m < 50$ eV\ \ ($m \simeq 40$ eV)\\
\hline
$\tau = 1$s; $m_{\nu_\mu} = 0,\, m_{\nu_\tau} = m$ & $0 \le m < 35$ eV 
 & $0 < m < 45$ eV\ \ ($m \simeq 30$ eV)\\
%\hline\hline
\end{tabular}
\end{table}

\begin{table}[h]
\caption{The results of the $\langle t \rangle$ analysis for different
cases.  The decay constant of the exponential luminosity is denoted by
$\tau$.  If the masses are zero, the most probable $\langle t
\rangle_S - \langle t \rangle_R = 0$.  For this $\langle t \rangle_S -
\langle t \rangle_R$, the allowed mass ranges are given in the second
column; the lower limit of zero is the most probable mass, and the
upper limit is excluded at the 90\% confidence level.  The smallest
value of $\langle t \rangle_S - \langle t \rangle_R$ not compatible
with $m = 0$ is $\langle t \rangle_S - \langle t \rangle_R = 0.09$ s.
The corresponding allowed mass ranges are given in the third column;
both the upper and lower limits are excluded at the 90\% confidence
level.  The most probable mass is given in parentheses.  For the third
case, because of the reduced width of the pulse, 0.03 s is used
instead of 0.09 s.}
\begin{tabular}{|l|l|l|}
\hline
case & result for $\langle t \rangle_S - \langle t \rangle_R = 0$ &
result for $\langle t \rangle_S - \langle t \rangle_R = 0.09$ s\\
\hline\hline
$\tau = 3$s; $m_{\nu_\mu} = 0,\, m_{\nu_\tau} = m$ & $0 \le m < 45$ eV 
 & $0 < m < 70$ eV\ \ ($m \simeq 45$ eV)\\
\hline
$\tau = 3$s; $m_{\nu_\mu} = m_{\nu_\tau} = m$ & $0 \le m < 35$ eV 
 & $0 < m < 45$ eV\ \ ($m \simeq 35$ eV)\\
\hline
$\tau = 1$s; $m_{\nu_\mu} = 0,\, m_{\nu_\tau} = m$ & $0 \le m < 25$ eV 
 & $0 < m < 40$ eV\ \ ($m \simeq 25$ eV)\\
%\hline\hline
\end{tabular}
\end{table}

\narrowtext
\noindent
to ask if there was evidence
for a nonzero mass, the evidence being a large $\chi^2$.  Strictly
speaking, if there was such evidence the mass would not be determined
with that test; one would reformulate the Reference to include a mass
and would define a new $\chi^2$, which would be minimized with respect
to the mass.  Nevertheless, our formulation works reasonably for small
masses.  Finally, because of its greater convenience in use and
interpretation, as well as its greater sensitivity, we advocate the
$\langle t \rangle$ technique.

We also considered the case in which both the $\nu_\mu$ and the
$\nu_\tau$ are massive.  For convenience, we took $m_{\nu_\mu} =
m_{\nu_\tau}$.  (Since the time delay is quadratic in the mass, there
is little difference from the one-mass case unless the masses are
similar.)  The results of the $\chi^2$ analysis are shown in Fig.~4,
and the results of the $\langle t \rangle$ analysis are shown in
Fig.~5.  As expected, with a better proportion of massive events in
the Signal, lower masses can be probed.  All of the results are
summarized in Tables II and III.

%%%%%%%%%%%%%%%%%%%%%%%%%%%%%%%%%%%%%%%%%%%%%%%%%%%%%%%%%%%%%%%%%%%%%%%%%%%%

\subsection{Comparison to previous work}

Various techniques for determining or limiting the $\nu_\mu$ and
$\nu_\tau$ masses from observations of supernova neutrinos have been
proposed.  Any such technique must be based on a neutral-current
signal and by necessity will contain events from other reactions with
similar signatures but caused by $\nu_e$ or $\bar{\nu}_e$. Also, in
neutral current events, one cannot determine the initial neutrino
energy on the event by event basis.  (In the neutrino-electron
scattering that is possible in principle, but not in practice).
Hence, one

\vbox{\vspace{10cm}}
\noindent
cannot directly determine the energy spectra of the
incoming $\nu_x$ neutrinos.

The most developed technique uses the signal from neutrino-electron
scattering in SK.  All flavors participate in this reaction, which has
no threshold.  Even though the $\nu_\mu$ and $\nu_\tau$ energies are
higher, their thermally-averaged cross sections are smaller than for
$\nu_e$ and $\bar{\nu}_e$ (which also have a charged-current channel).
Thus massless events are necessarily part of the irreducible
background.  There are also isotropic background events from the
copious $\bar{\nu}_e + p \rightarrow e^+ + n$ reaction; by considering
only events in the forward cone of half-angle about 25 degrees
(determined by the angular resolution of the \v{C}erenkov detector),
one can eliminate about 95\% of the isotropic
background\cite{Krauss,Fiorentini}.  From Table I it follows that if
just the $\nu_\tau$ is massive, in the forward cone there are about
700 $m = 0$ events and about 60 $m > 0$ events.  The test for a mass
is to check whether the events in the forward cone fall off more
slowly in time than those outside the cone.  Since the number of
massive events is small, one has to look for a large delay.  At such
late times in the tails of the scattering rates, the time-independent
background rate is not at all negligible.

The most detailed analysis of the neutrino-electron scattering case
was given in Ref.~\cite{Krauss}.  The statistical test for a mass was
done by a complicated likelihood matching scheme, and sensitivity to a
mass of about 50 eV was found.  Another detailed analysis was given in
Ref.~\cite{Fiorentini}.  The statistical test for a mass was simple,
and was based on looking for an excess of events at late times, where
an excess was defined as three times the Poisson error.  In this case,
sensitivity to a mass of only about 150 eV was found.  The question
arises if this poorer limit was caused by the less sophisticated
statistical technique.  Interestingly, it is not.  It is pointed out
in Ref~\cite{Fiorentini} that the authors of Ref.~\cite{Krauss} use a
luminosity which decays roughly exponentially, with a time constant of
$\tau = 1$ s (in contrast to the time constant of $\tau = 3$ s used in
this work).  Such a sharp time distribution makes distinguishing the
effects of a mass much easier.  In Ref.~\cite{Fiorentini}, it is shown
that using such a quickly-decaying luminosity and the same simple
statistical technique that sensitivity to 50 eV can also be obtained.

For comparison, we set the exponential time constant in the luminosity
to $\tau = 1$ s and repeated our analysis.  For $t_{max} = 3$ s and a
bin size of 0.5 s, we found sensitivity to about 25 eV in the one-mass
case.  The results are also presented in Tables II and III.  The
advantages of the method discussed here, as demonstrated by this
comparison, are the larger number of events with mass and the lower
proportion of massless events with a similar signature.

\begin{figure}[t]
\epsfxsize=3.25in \epsfbox{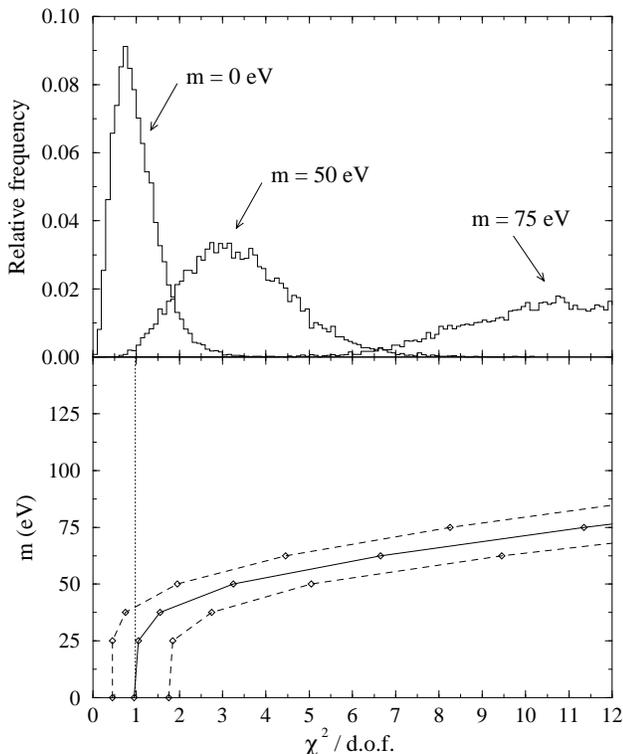}
\caption{The results of the $\chi^2$ analysis for a massive $\nu_\mu$
and $\nu_\tau$, taken to have the same mass.  The figure is otherwise
the same as Fig. 2.}
\end{figure}

%%%%%%%%%%%%%%%%%%%%%%%%%%%%%%%%%%%%%%%%%%%%%%%%%%%%%%%%%%%%%%%%%%%%%%%%%%%%
%%%%%%%%%%%%%%%%%%%%%%%%%%%%%%%%%%%%%%%%%%%%%%%%%%%%%%%%%%%%%%%%%%%%%%%%%%%%

\section{Intermediate-mass and high-mass cases}
 
For a low or zero mass ($m_{\nu_\tau} \lesssim 150$ eV), the effects
on the signal $S(t)$ are minimal, and the time-independent background
is negligible.  For an intermediate mass ($150 {\rm\ eV} \lesssim
m_{\nu_\tau} \lesssim 1$ keV), the effects on $S(t)$ are substantial,
and the massive component of $S(t)$ will be well-separated from the
the massless component.  The time-independent background does not have
a large effect, but would have to be taken into account.  For a large
mass ($m_{\nu_\tau} \gtrsim 1$ keV), the massive component of $S(t)$
is so delayed and dispersed that its rate is comparable to or below
the time-independent background rate.  Given the actual data, one can
immediately determine which of these cases applies.  There are
analysis techniques that are optimal for each case.  It is ``fair'' to
determine the choice of technique from the crude characteristics of
the data.

\begin{figure}[t]
\epsfxsize=3.25in \epsfbox{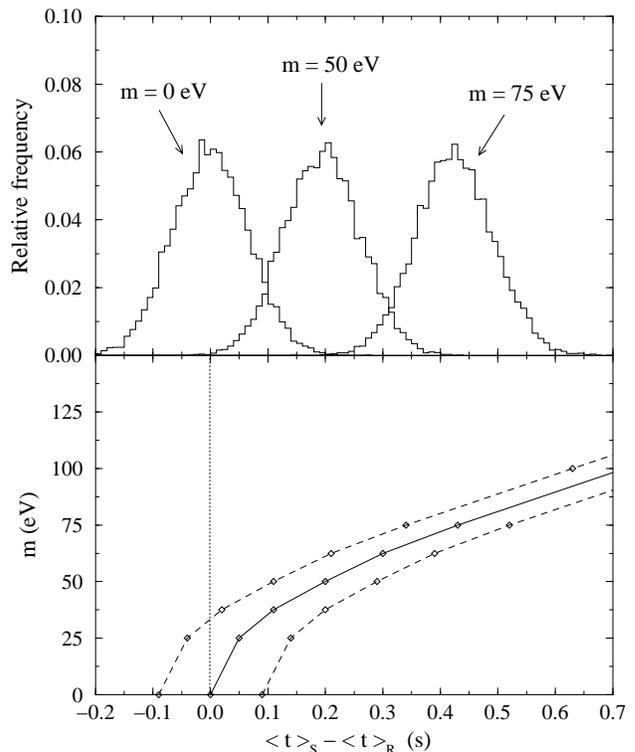}
\caption{The results of the $\langle t \rangle$ analysis for a massive
$\nu_\mu$ and $\nu_\tau$, taken to have the same mass.  The figure is
otherwise the same as Fig. 3.}
\end{figure}

The intermediate-mass case would be rather easy to handle.  The value
of $t_{max}$ would have to be increased and the time-independent
background rate included.  The $\chi^2$ analysis above was designed to
test whether or not a mass was necessary to explain the data.  For a
small mass, it can be used to determine that mass.  As noted earlier,
for a large and obvious mass, it would be better to revise the
$\chi^2$ analysis so that the Reference $R(t)$ was that appropriate
for a given mass.  Then the $\chi^2$ could be minimized to find the
unknown mass and its error.  The $\langle t \rangle$ technique
requires only the changes noted.  For such a large mass, the
dispersion (broadening) also becomes a useful measure of the mass.
Almost any technique would work in this case since the signal would be
so obvious.

The large-mass case, like the low-mass case, is again a marginal
analysis since we are by definition looking at the limit of
detectability.  For a large mass, the delays are large compared to the
width of the pulse at the source, and the integral in
Eq.~(\ref{eq:rate}) can be evaluated by assuming that the time
distribution of the initial pulse is a delta function.  The scattering
rate (per s) is
\begin{equation}
\frac{dN_{sc}}{dt} =
\frac{C}{2 t}\, \widetilde{E} f(\widetilde{E})
\left[\frac{\sigma(\widetilde{E})}{10^{-42} {\rm cm}^2}\right]\,,
\end{equation}
where $C$ is defined in Eq.~(\ref{eq:C}), and $\widetilde{E}$ in MeV
is defined as $\widetilde{E} = m \sqrt{0.515 D/t}$ (see
Eq.~(\ref{eq:delay})), with $m$ in eV, $D$ in 10 kpc, and $t$ in s.
Note also that $f$ is in MeV$^{-1}$.  The time $t$ is measured from
the arrival of the $\bar{\nu}_e$ events.  For $m = 1$ keV, the signal
is still several times the time-independent background for hundreds of
seconds.  As the mass increases, the height of the signal rate falls
very quickly.

Even if $S(t) < B$ at all times, where $B$ is the time-independent
background rate, it is still possible to determine a mass by looking
for an excess of counts in some long time interval.  We assume that
the expected number of signal events is present (the Poisson
fluctuation of the signal number will turn out to be a small effect).
This analysis is therefore more model-dependent than the low-mass
case, since the number of events enters directly, rather than only
through the fluctuations.  While $t$ is defined by the arrival of the
$\bar{\nu}_e$ events, they are obviously not included in the counts
for this analysis.  Only a finite range of neutrino energies
contribute significantly, and the largest energy is of order 5 times
the smallest.  The largest delay will thus be of order 25 times the
smallest.  In this case, the simplest and most model-independent thing
to do is to begin the counting at $t = 0$.

We assume that the background rate $B$ is well-known.  The end of the
counting interval $t_{max}$ is to be determined.  The requirement of a
statistically significant excess of counts is $N_B + N_S > N_B + n
\sqrt{N_B}$, where $n$ is the number of sigmas (the number of counts
is large enough to treat the Poisson distribution as a Gaussian.)  Any
large excess in the number of events will be wholly attributed to the
signal events, of which there are $N_S$ expected.  Using $N_B = B
t_{max}$, this can be rewritten as $t_{max} < N_S^2/n^2 B$.  Note that
this is independent of mass.  The requirements for $t_{max}$ are:
\begin{equation}
{\rm signal\ width} < t_{max} < \frac{N_S^2}{n^2 B}\,.
\end{equation}
If the interval is not as wide as the signal, signal events will be
lost.  If it is wider than the signal, too many background events will
be included.  The largest possible mass that can be seen with this
technique is the one for which the signal width is as wide as the
right-hand side of the equation above.  This is
\begin{equation}
m_{max} = E_{min} \frac{N_S}{n\sqrt{0.515 D B}}\,,
\end{equation}
where $m$ is in eV, $E_{min}$ is in MeV, $D$ is in 10 kpc, and $B$ is
in s$^{-1}$.  For the $\nu_x$ excitation of $^{16}$O, we take $E_{min}
= 25$ MeV; below that energy, the product $f(E)\sigma(E)$ is
essentially zero.  In order to reduce the time-independent background
rate, we use only the inner 22.5 kton volume for this large-mass test,
which reduces the number of $\nu_\tau$ signal events to $N_S = 250$.
For this volume, the background rate has been measured\cite{SKbg} to
be of order 0.1 s$^{-1}$.  At the three-sigma level ($n = 3$), the
maximum detectable mass is then about $m_{max} = 9$ keV.  For this
mass, even the peak of the signal rate is a factor several below the
time-independent background rate.  Also for this mass, $t_{max}$ is of
order $10^5$ s, so the Poisson error on the number of background
events is at the 1\% level.  We have assumed that the error on the
background rate $B$ is not larger than that.  The above analysis is
optimized for a flat signal.  However, the signal is actually peaked
at a time smaller than $t_{max}$, and by increasing $E_{min}$ to 42
MeV, one still includes about 90\% of the signal events.  While more
model-dependent, this increases the maximum detectable mass to about
$m_{max} = 14$ keV.

For comparison, we estimate how large $m_{max}$ would be if the signal
from neutrino-electron scattering were used.  Since the signal is
forward-peaked, the background can be substantially reduced with an
angular cut.  In this case, it makes sense to use the entire 32 kton
volume.  If 95\% of the background can be removed, and the
time-independent background rate of 0.1 s$^{-1}$ used above for the
inner 22.5 kton can be used for the full volume, then $B \approx
0.005$ s$^{-1}$.  Assuming that no signal events are lost with this
cut, the number of events for $E_{min} = 5$ MeV is about $N_S = 60$.
At the three-sigma level, the maximum detectable mass is about
$m_{max} = 2$ keV, comparable to the estimate in Ref.~\cite{Krauss}.

If the $\nu_\tau$ events appear to be missing, a large-mass search as
above can be made.  If nothing is found, there are three
possibilities.  The first possibility is that the mass is greater than
10 keV or so, and that it is stable over the time it takes to travel
from the supernova, about $3 \times 10^{4}$ years.  Then its signal is
so dispersed that it cannot distinguished against the background.
However, as pointed out in Ref.~\cite{Wolfenstein}, any neutrino with
a mass greater than 10 keV or so would likely decay in such a time
(this avoids violation of the cosmological bound on the neutrino
masses, see Ref.~\cite{Raffelt} and references therein).  The second
possibility is that the mass was large enough that the neutrinos
decayed, and that their decay products were not detected.  The third
possibility is that the $\nu_\tau$ neutrino was not produced in the
supernova, or at least significantly differently than expected.  For
example, if the $\nu_\tau$ temperature were much lower than 8 MeV,
there would be essentially no $\nu_\tau$ events detected.  These three
possibilities cannot be distinguished without additional evidence.

%%%%%%%%%%%%%%%%%%%%%%%%%%%%%%%%%%%%%%%%%%%%%%%%%%%%%%%%%%%%%%%%%%%%%%%%%%%%
%%%%%%%%%%%%%%%%%%%%%%%%%%%%%%%%%%%%%%%%%%%%%%%%%%%%%%%%%%%%%%%%%%%%%%%%%%%%

\section{Conclusions and discussion}

One of the key points of our technique is that the abundant
$\bar{\nu}_e$ events can be used to calibrate the neutrino luminosity
of the supernova and to define a clock by which to measure the delay
of the $\nu_x$ neutrinos.  The internal calibration very substantially
reduces the model dependence of our results.  The measurement of time
relative to the $\bar{\nu}_e$ signal allows us to be sensitive to rather
low masses.  Without such a clock, one cannot determine a mass limit
with the $\langle t \rangle$ technique advocated here, since the
absolute delay is unknown.  Instead, one would have to constrain the
mass from the observed dispersion of the events.  Our calculations
indicate that while a significant delay can be seen for $m = 50$ eV,
the dispersion does not become significant until $m = 150$ eV or
greater.

We first assumed that one of $\nu_\mu$ and $\nu_\tau$ masses was
nonzero, and the other negligibly small.  For convenience, we referred
to the heavier one as $\nu_\tau$, though it is impossible to tell the
difference.  The results are given in Figs.~2 and 3.  If it were known
that the masses were almost degenerate, than a stricter limit can
be placed.  Those results are given in Figs.~4 and 5.  If nothing more
is known, the most conservative thing to do is to take the one-mass
limit for each of $\nu_\tau$ and $\nu_\mu$.  As shown in Table III, if
no statistically significant difference of the Reference and Signal is
seen one can put an upper limit of 45 eV if one assumes that only one
mass in nonvanishing, and 35 eV if one assumes that both $\nu_{\mu}$
and $\nu_{\tau}$ are massive (and that the masses are the same).

Given the large statistics of the $\nu_x$ signal used here, one might
wonder why the time delay is not larger and the mass sensitivity is
not lower than we report here.  The $\nu_x$ average energy is about 25
MeV.  For $E \approx 25$ MeV, $m = 50$ eV and $D = 10$ kpc, the delay
is about 2 s.  However, from Eq.~(\ref{eq:rate}), what matters for the
event rate is the peak of the product $f(E)\sigma(E)$.  Since the
cross section for the $^{16}$O excitation is very steep in energy, the
peak energy is large, about 60 MeV.  For $E \approx 60$ MeV, $m = 50$
eV and $D = 10$ kpc, the delay is about 0.4 s.  In both cases, these
delays are for about 1/3 of the events in the Signal, so for a large
integration time $t_{max}$ the difference $\langle t \rangle_S -
\langle t \rangle_R$ would be about 1/3 of these delays.  For moderate
$t_{max}$, as used in the main analysis, the shift is slightly smaller
(though more significant than for a larger $t_{max}$).

These considerations show that the delay is reduced, and the
statistical significance decreased, by the seemingly irreducible
background of the $\bar{\nu}_e$ events at low energies as well as by
the background caused by the massless $\nu_{\mu}$.  Besides, since the
energy of the outgoing neutrino cannot be measured (or even the
excitation energy in $^{16}$O), it is not possible to measure the
energy spectrum of the $\nu_\tau$ neutrinos.  Thus the $\nu_x$
temperature can only be constrained from the total number of events.

The situation can be contrasted with the $\bar{\nu}_e$ mass limit of
about 20 eV from SN1987A established with only a handful of events and
no independent clock.  There, however, it was possible to determine
the incoming neutrino energy on the event by event basis, and to
compare the neutrino energies versus time to the theoretical
expectation.  Moreover, the SN1987A was at about 50 kpc, compared to
the 10 kpc assumed for the next Galactic supernova, and a lower
typical energy should be used in the delay formula of the detected
$\bar{\nu}_e$ events than for the $\nu_x$ neutral current scattering
on $^{16}$O.

Some of the important parameters used here are not well known, though
were treated as such.  However, once there is actually a supernova,
the model uncertainties will be greatly reduced by the $\bar{\nu}_e$
data.  For example, the binding energy $E_B$ and the $\bar{\nu}_e$
temperature will be determined.  Other questions that can be resolved
include the time dependence of the temperature, and whether a
one-parameter thermal spectrum is sufficient to describe the energy
spectra.  Once a supernova is observed, the technique presented here
can easily be run with the new parameters or necessary modifications.
Second, for small changes in some parameters, the mass sensitivity
does not change much.  Note that this is especially true if mass is
unrecognizably small, and we are making a limit.

The results of this paper are valid for either Dirac or Majorana
neutrinos.  We only considered stable neutrinos.  The effects of
decaying neutrinos on mass limits from supernovae are discussed in
Ref~\cite{Wolfenstein}.  We also considered unmixed neutrinos.  Vacuum
oscillations among $\nu_\tau$, $\nu_\mu$, and their antiparticles are
irrelevant since the numbers of neutrinos of each flavor are assumed
to be equal.  Vacuum oscillations between $\nu_\tau$ and $\nu_e$ or
$\nu_\mu$ and $\nu_e$ and their antiparticles should have an
observable effect on the $\bar{\nu}_e$ spectrum.  Oscillations to
sterile neutrinos would also have an effect.  The effects of either
vacuum or matter-enhanced neutrino mixing on the neutrino signals are
considered in, e.g., Ref.~\cite{Qian}.

In conclusion: We have presented a rather general method, including a
thorough statistical analysis, of extracting information about the
possible $\nu_\tau$ and $\nu_\mu$ masses from the future detection of
a Galactic supernova neutrino burst by the SuperKamiokande detector.
When such an event in fact occurs, the existing mass limits will be
vastly improved and will approach, or cross over, the cosmological
bound.

%%%%%%%%%%%%%%%%%%%%%%%%%%%%%%%%%%%%%%%%%%%%%%%%%%%%%%%%%%%%%%%%%%%%%%%%%%%%
%%%%%%%%%%%%%%%%%%%%%%%%%%%%%%%%%%%%%%%%%%%%%%%%%%%%%%%%%%%%%%%%%%%%%%%%%%%%

\section*{ACKNOWLEDGMENTS}
This work was supported in part by the US Department of Energy under
Grant No. DE-FG03-88ER-40397.  J.F.B. was supported by a Sherman
Fairchild fellowship from Caltech.  We thank H.A. Bethe, B.W.
Filippone, and Y.-Z. Qian for discussions.

%%%%%%%%%%%%%%%%%%%%%%%%%%%%%%%%%%%%%%%%%%%%%%%%%%%%%%%%%%%%%%%%%%%%%%%%%%%%
%%%%%%%%%%%%%%%%%%%%%%%%%%%%%%%%%%%%%%%%%%%%%%%%%%%%%%%%%%%%%%%%%%%%%%%%%%%%

%%%%%%%%%%%%%%%%%%%%%%%%%%%%%%%%%%%%%%%%%%%%%%%%%%%%%%%%%%%%%%%%%%%%%%%%%%%%
%%%%%%%%%%%%%%%%%%%%%%%%%%%%%%%%%%%%%%%%%%%%%%%%%%%%%%%%%%%%%%%%%%%%%%%%%%%%

\end{document}